\documentclass{PoS}

\title{Recommendations for PDF usage in LHC predictions}

\ShortTitle{Recommendations for PDF usage in LHC predictions}

\author{\speaker{Ringail\. e Pla\v cakyt\. e}\\
        Universit\" at Hamburg, II. Institut f\" ur Theoretische Physik \\
        E-mail: \email{ringaile@mail.desy.de}} 

\abstract{
    
    A short review of the currently available modern parton distribution functions (PDFs) 
    and the theory predictions obtained using those PDFs for several benchmark processes
    at LHC, including Higgs boson production, is presented in this write-up. 
    It includes the discussion on theory assumptions made in the determination procedure 
    of PDFs and an impact 
    on the differences in the obtained predictions, followed
    by the alternative to PDF4LHC recommendations for the usage of PDF sets for theory 
    predictions at the LHC.
}

\FullConference{XXIV International Workshop on Deep-Inelastic Scattering and Related Subjects\\
		11-15 April, 2016\\
		DESY Hamburg, Germany}

\begin{document}

\section{Overview of modern PDFs}

The wealth of precise measurements from the Large Hadron Collider (LHC) including 
the new data at the highest center-of-mass collision energies ever achieved ($\sqrt s=  13$ TeV)
are used to search for physics phenomena beyond the SM (BSM). 
The measurements are compared to precise theoretical predictions, which, ideally, 
incorporate higher order radiative corrections in Quantum Chromodynamics (QCD) and the electroweak sector of the SM.
Parton distribution functions (PDFs) in the proton are an essential input for any cross
section prediction at hadron colliders. 
Despite steady improvements in the accuracy of PDF determinations in the last years, 
the uncertainties associated with PDFs, likewise and the strong coupling $\alpha_s(M_{Z})$ 
and heavy quark masses, still dominate many calculations of cross sections for SM processes at
the LHC.
The currently available PDF sets are CJ15~\cite{r1}, accurate to NLO in QCD, as well as ABM12~\cite{r2},
CT14~\cite{r3}, HERAPDF2.0~\cite{r4}, JR14~\cite{r5}, MMHT14~\cite{r6}, and NNPDF3.0~\cite{r7} to NNLO in QCD.
Both theoretical and experimental inputs used in the PDF analyses have direct impact on the
obtained nonperturbative parameters, namely, the fitted PDFs, the value of $\alpha_s(M_{Z})$
and the quark masses.  For precision predictions in Run 2 of the LHC it is therefore
very important to quantify those effects
in order to reduce the uncertainties in the nonperturbative input parameters.
The benchmark of individual PDFs, including the combined sets as proposed in the recent 
PDF4LHC recommendations~\cite{r8}, is performed for various processes at hadron
colliders and presented together with the
detailed discussion on the underlying theory assumptions in the PDF analysis procedure,
followed by the alternative recommendations for the usage of PDFs for theory predictions at
the LHC.
%

\section{Cross section predictions for the LHC}

The detailed overview of the currently available data which can be used to determine PDFs
together with the PDF fit results from different groups is provided in the original study in Ref.~\cite{r9}.
The basic theoretical issues for a consistent determination of the PDFs from DIS and other 
hard scattering data is described in the same reference.
In following, few selected examples are discussed, namely the value of the strong coupling constant,
Higgs cross section predictions at LHC, hadro-production of heavy quarks and heavy $W'$ boson
production. 

\subsection{The strong coupling constant}

The value of the strong coupling constant $\alpha_s(M_Z)$ 
is one of the most important parameters in QCD which has a direct impact on the
size of a number of cross sections at the LHC such as Higgs production.
In the context of global PDF analyses $\alpha_s(M_Z)$ has a particular interest 
due to its correlation with the gluon PDF and the charm-quark mass.
The value of $\alpha_s(M_Z)$ is determined from a large number of different processes and 
methods at different scales and yields the world average 0.1181 $\pm$ 0.0013 at NNLO~\cite{r14}.
While the world average of $\alpha_s(M_Z)$ is closely related 
to the values determined in PDF fits, two important aspects specific to PDF analyses
are addressed here: 
First, in some PDF analyses
$\alpha_s(M_Z)$ is treated as a free parameter which allows to control its correlation 
with other PDF parameters. Second, a large spread of $\alpha_s(M_Z)$ values 
are used (or determined) in PDFs, ranging from 0.1132 to 0.1183, see Table~\ref{as_tab}.
These differences can be traced back to different data sets used or to different
theory assumptions applied, as indicated in Tab~\ref{as_tab}.
One example is the data for the hadro-production of jets at LHC where full higher than 
NLO order corrections are not yet available but are known to be significant - various
PDF groups employ different methods to account for this inconsistency in NNLO fits
which in own turn, may lead to different $\alpha_s(M_Z)$ values. Examples
of different methodology in PDF fits which are potential sources of discrepancies
in $\alpha_s(M_Z)$ values are nuclear corrections, heavy-flavour schemes, etc. as
discussed in Ref.~\cite{r9}.
In addition, the bias also may be introduced by fixing $\alpha_s(M_Z)$ 
to the same value at different orders (NLO and NNLO) in PDF sets used in 
the combination procedure of PDF4LHC recommendations~\cite{r8}, which is in contradiction
with the precision determinations of $\alpha_s(M_Z)$ at different orders in perturbation
theory.

\begin{table}[!ht]
 \centering
 \small
  \begin{tabular}{|c|c|c|}
    \hline \hline
    ABM12~\cite{r2}  & 0.11320 $\pm$ 0.0011 & fit at NNLO \\
    \hline
    CJ15~\cite{r1} & 0.11830 $\pm$ 0.0002 & fit at NLO \\
    \hline
    CT14~\cite{r3} & 0.118&  assumed at NNLO \\
    \hline
    HERAPDF2.0Jets~\cite{r4} & 0.1183 $+$0.0040 $-$0.0034 & fit at NLO \\
    \hline
    JR14~\cite{r5}  & 0.11360$\pm$0.0004 & dynamical fit at NNLO  \\
              & 0.11620$\pm$0.0006 & standard fit at NNLO \\ 
    \hline
    MMHT14~\cite{r9} & 0.118 &  assumed at NNLO \\
    \hline
    NNPDF3.0~\cite{r7} &  0.115 $-$0.121 & assumed at NNLO; preferred value 0.118 \\
    \hline
    PDF4LHC15~\cite{r8} & 0.118 & assumed at NNLO \\
    \hline \hline
  \end{tabular}
  \caption{Values of $\alpha_s(M_Z)$ obtained or used in the nominal PDF sets of the various groups.}
 \label{as_tab} 
\end{table} 

\subsection{Higgs cross section predictions at LHC}

The dominant production mechanism for the SM Higgs boson at the LHC is the gluon-gluon
fusion process which is known to N$^3$LO in QCD~\cite{r10,r11}. Currently the largest source 
of uncertainties in the predictions of the Higgs cross section are the value of $\alpha_s$ and the PDFs. \\
A large spread for Higgs predictions observed from different PDFs with a range 
$ 38.0 - 42.6$~pb using the nominal value of $\alpha_s(M_Z)$ ($11\%$) and 
$39.0 - 44.7$~pb if a fixed value of $\alpha_s(M_Z)$ = $0.115- 0.118$ ($13\%$) is 
used (for details please see the original study~\cite{r9}). This illustrates the importance 
of controlling the accuracy and the correlation of the strong coupling constant with the
PDF parameters and the heavy quark masses in fits.
As an illustration, Tab.~\ref{h_tab} shows the values of the charm-quark mass $m_c^{pole}$, $\alpha_s(M_Z)$ 
and $\chi^2$ value obtained using the open-source package \texttt{xFitter}~\cite{rxf} 
for the HERA charm data~\cite{r12} together with the calculated Higgs cross section at NNLO, using MSTW PDFs.
Here, a linear rise of the cross section for 
increasing values of charm mass is observed (3$\%$ for the best fit $\alpha_s(M_Z)$
in the range from 1.15 to 1.55 GeV 
and 1$\%$ for fixed). The latter case leads to a best fit of $m_c^{pole}$~=~1.2~GeV
which is significantly smaller than the nominal fit with $m_c^{pole}$~=~1.4~GeV.
These values are not compatible with the world average of the PDG~\cite{r13} and indicate
that the charm-quark mass effectively takes over the role of a 'tuning' parameter.
Similar results have been observed in other PDF analyses (please see
original study~\cite{r9}). These results are in agreement with the results shown in e.g. CT10 study~\cite{r14},
where about 2$\%$ difference in Higgs cross section is observed for $m_c^{pole}$ 
changing in the range from 1.0 to 1.36 GeV (please note the smaller value of the $m_c^{pole}$ range
than the one considered above).
\begin{table}[!ht]
 \centering
 \small
  \begin{tabular}{|c|c|c|c|c|}
    \hline \hline
    $m_c^{pole} $  & $\alpha_s(M_Z)$ & $\chi^2/NDP$ & $\sigma(H)^{NNLO} [pb]$ best fit $\alpha_s(M_Z)$ &  $\sigma(H)^{NNLO} [pb]$ $\alpha_s(M_Z)$ = 0.118 \\
    \hline
    1.15 & 0.1164&  78/52 (71/52) & 41.01 & (42.05) \\
    \hline
    1.2 & 0.1166 & 76/52 (70/52) & 41.18 & (42.11) \\
    \hline
    1.25 & 0.1167 & 75/52 (76/52) & 41.33 &(42.17) \\
    \hline
    1.3 & 0.1169 & 76/52 (77/52) &  41.48 & (42.25) \\ 
    \hline
    1.35 & 0.1171 & 78/52 (79/52) & 41.68 &(42.30) \\
    \hline
    1.4 & 0.1172 & 82/52 (83/52) & 41.83 &(42.36) \\
    \hline
    1.45 & 0.1173 & 88/52 (89/52) & 42.00 &(42.45) \\
    \hline  
    1.5 & 0.1173 & 96/52 (96/52) & 42.14 & (42.51) \\
    \hline
    1.55 & 0.1175 & 105/52 (106/52) & 42.29 & (42.58) \\
    \hline \hline
  \end{tabular}
  \caption{The values of the charm-quark mass and $\alpha_s(M_Z)$ in MMHT analysis~\cite{r9} 
  together with the value for $\chi^2/NDP$ for the HERA data~\cite{r12} and the Higgs 
  cross section at NNLO at $\sqrt s$ = 13 TeV for $m_H$ = 125 GeV at the 
  nominal scale $m_H$. The numbers in parentheses are obtained with the value of $\alpha_s(M_Z)$ 
  fixed to 0.118}
 \label{h_tab} 
\end{table} 
\\
Finally, for the calculation of PDF uncertainties in precision observables (such as the
Higgs cross section), the combined PDF set is recommended to use in PDF4LHC recommendations~\cite{r8}.
It is unclear why PDFs should be treated differently in this case than in the case 
of comparisons between data and theory for SM measurements (point 1 of
recommendations), since these differences are smeared out in the combination and,
in addition, the comparison is limited to only three PDF sets.

\subsection{Hadro-production of heavy quarks}

\begin{figure}[t!]
\begin{center}
\includegraphics[width=0.82\textwidth, angle=0]{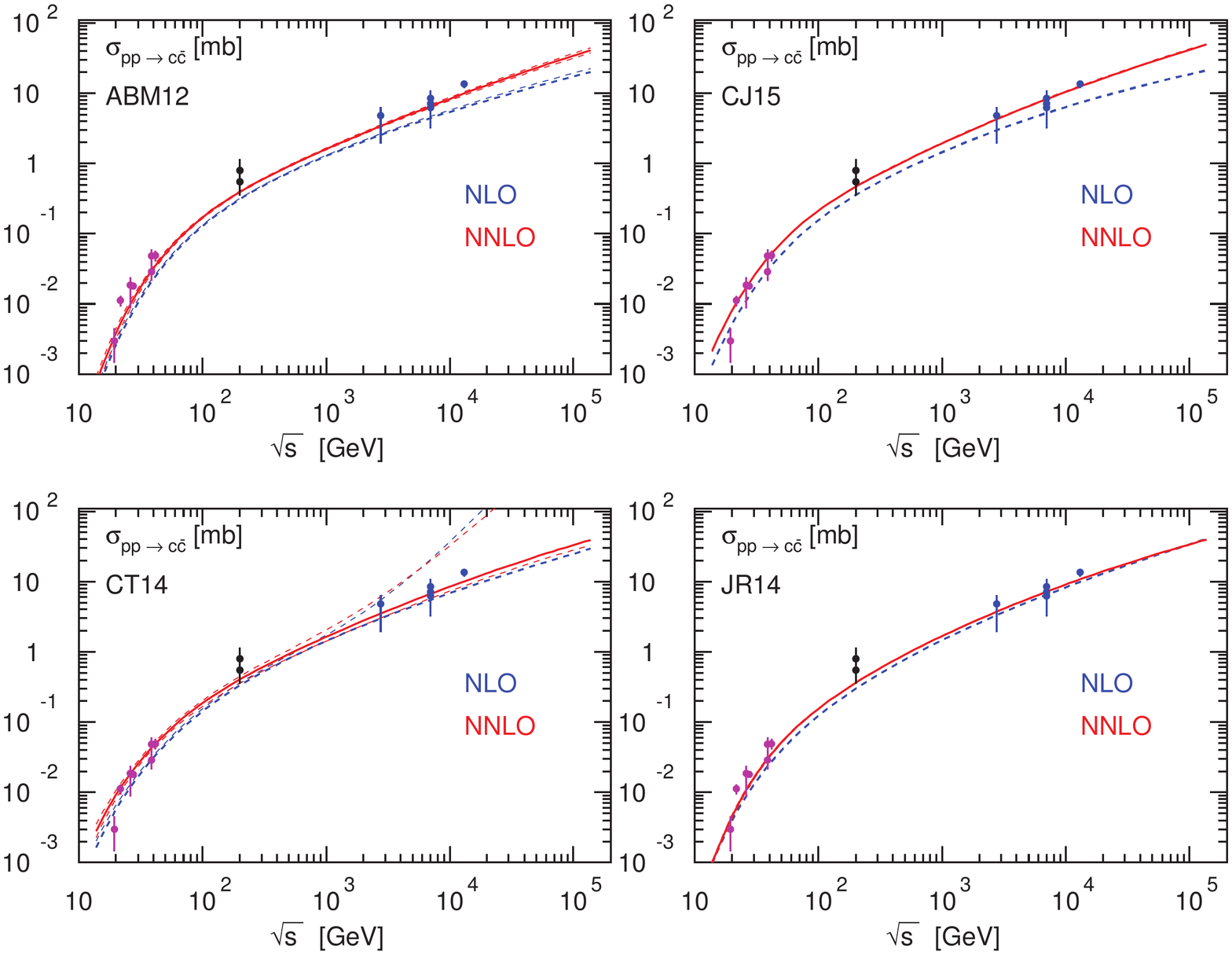}
\hspace*{3mm}
\includegraphics[width=0.80\textwidth, angle=0]{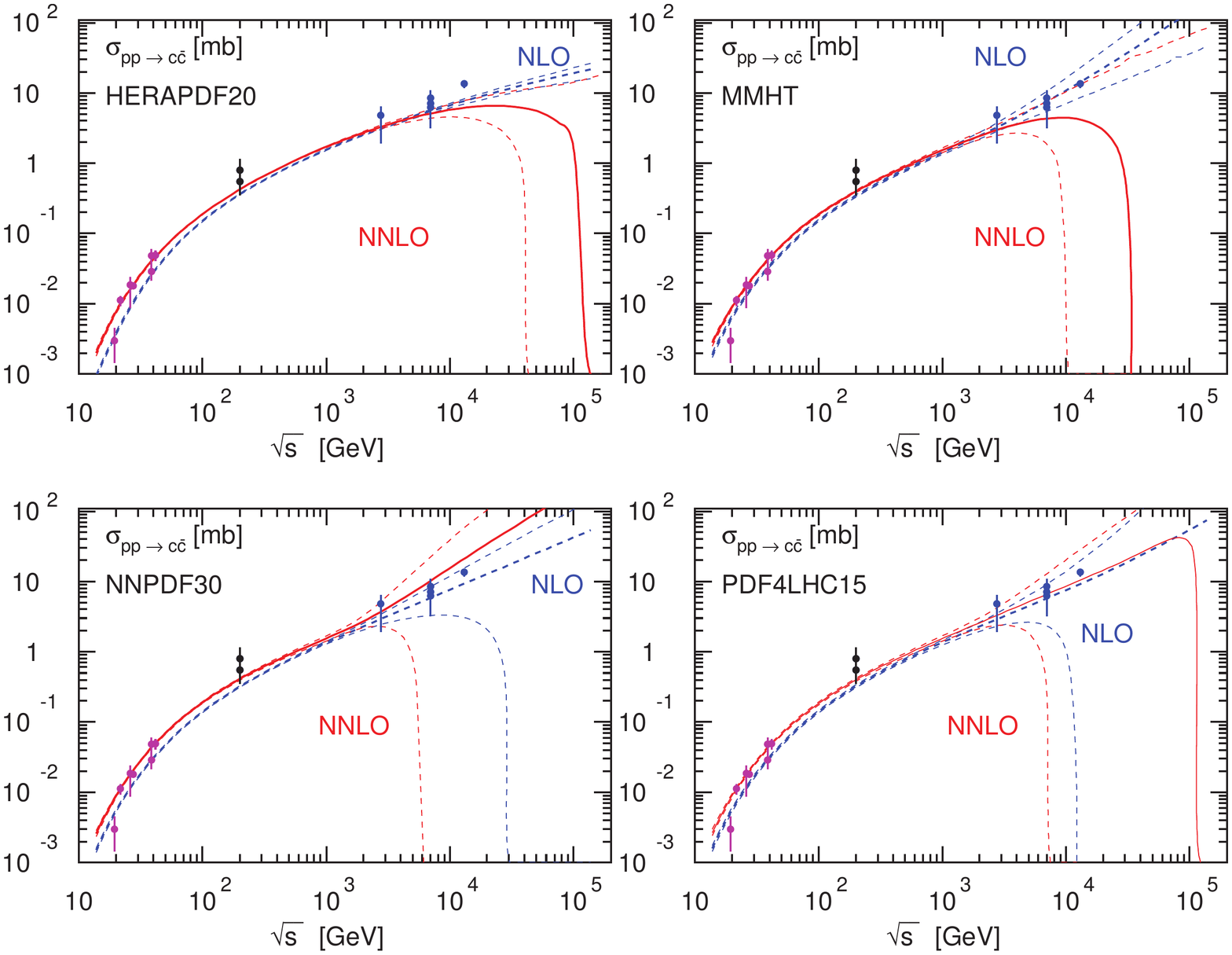}
\caption{\small Theoretical predictions for the total $pp \rightarrow c\bar c$ cross section 
    as a function of the center-of-mass energy $\sqrt{s}$ at NLO (dashed lines) and NNLO (solid lines) 
    in the $\overline MS$ scheme with $m_c(m_c)$ = 1.27 GeV and scale choice $\mu_R = \mu_F = 2m_c(m_c)$ 
    using various PDF sets (PDF uncertainties shown as dashed lines)}
\label{f_cc} 
\end{center}
\end{figure}

Charm-quark hadro-production provides the possibility to check the consistency of different PDFs.
The inclusive cross section of the reaction $pp \rightarrow c\bar c$ compared with theory 
predictions at NLO and NNLO using various PDFs as a function of the center-of mass
energy $\sqrt{s}$ to available experimental data is shown in Fig.~\ref{f_cc}.
All predictions agree well with data at low energies but start to behave differently for 
HERAPDF2.0, MMHT14 or NNPDF3.0 at energies above $\sqrt{s} \simeq \mathcal{O}$(10) TeV. 
Despite that at these energies the associated PDF uncertainties become very large,
the striking feature seen in Fig.~\ref{f_cc} is the negative cross sections for
HERAPDF2.0, MMHT14 and PDF4LHC15 at NNLO. This feature is caused by the negative gluon in the NNLO fit 
of those PDFs illustrating
an instability of the perturbative expansion of the inclusive $pp \rightarrow c\bar c$ cross
sections at energies when the contribution from the quark-gluon channel dominates.

\subsection{Heavy $W'$ boson}

\begin{figure}[t!]
\begin{minipage}[c]{0.62\textwidth}
\includegraphics[width=0.85\textwidth, angle=0]{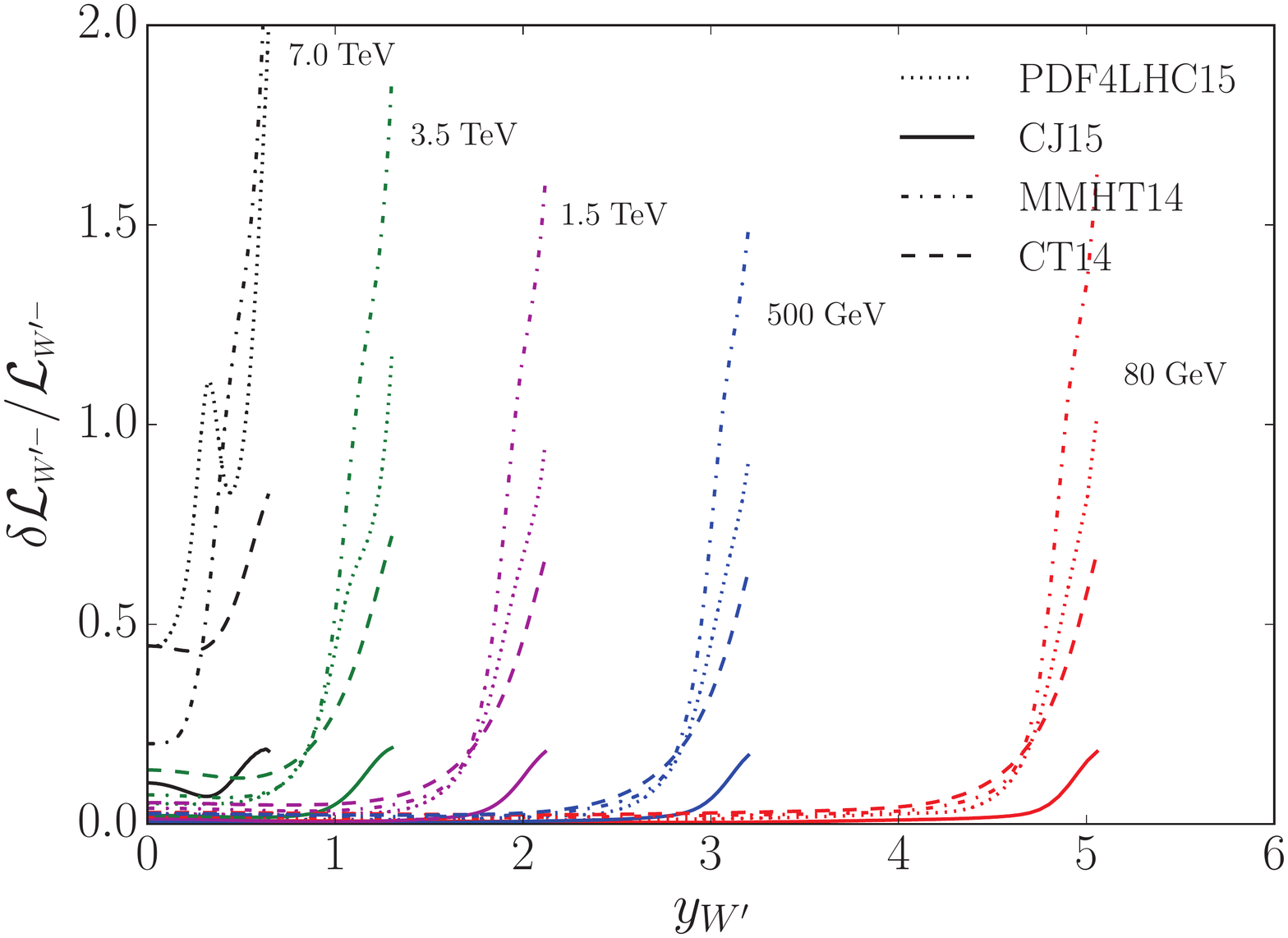}
  \end{minipage}\hfill
  \begin{minipage}[c]{0.36\textwidth}
\caption{\small Relative uncertainty $\delta \mathcal{L}_{W'^{-}}/\mathcal{L}_{W'^{-}}$ in the $W'^{-}$ 
    luminosity as a function of rapidity $y_{W'}$ for the
    combined PDF4LHC15 set (dotted), the CJ15 (solid), MMHT14 (dot-dashed), and CT14 (dashed) PDFs
    for various $W'$ masses. All PDF uncertainties have been scaled to a common
    68$\%$ c.l. }
\label{f_w} 
\end{minipage}
\end{figure}

The PDFs at large-$x$ are not known precisely due to limited available experimental data and
due to various approximations in the treatment of nuclear corrections for deuterium data, 
target mass and higher twist effects. This can be illustrated with the production of a heavy $W'$ 
boson as a function of the $W'$ rapidity, $y_{W'}$.  Fig.~\ref{f_w} shows the uncertainty in the 
parton luminosity for a produced negatively charged $W'$ boson for various PDF sets.
Here a very large range of uncertainties is observed for the various PDF sets due to different tolerance 
criteria and methodologies used for the treatment of data at high values of $x$.
The smallest uncertainty is obtained for
the CJ15 PDF set, which makes use of low invariant mass data to constrain the high-$x$ region, and
does not employ additional tolerance factors inflating the uncertainties. The MMHT and CT14
PDF sets have larger errors, due to stronger cuts on low-mass DIS data and larger tolerances, and
consequently the combined PDF4LHC15 set gives similarly large uncertainties.
This example illustrates the problematic nature of statistically combined PDF sets that have
been determined using very different theoretical treatments of the high-$x$ region, leading to a
possible overestimate of the uncertainties at these kinematics.

\section{Alternative recommendations for PDF usage}

Based on the presented benchmark results and the aspects discussed regarding the official PDF4LHC 
recommendations~\cite{r8}, we propose modifications to the recommendations for PDF usage 
at the LHC which allow to retain the predictive capability of the individual PDF sets. 
Two cases can be distinguished:

\begin{enumerate}
    \item[1] {\textbf{Precise theory predictions:}}
        Recommendation: Use the individual recent PDF sets, currently ABM12~\cite{r2}, CJ15~\cite{r1},
        CT14~\cite{r3}, JR14~\cite{r5}, HERAPDF2.0~\cite{r4}, MMHT14~\cite{r6}, and NNPDF3.0~\cite{r7} 
        (or as many
        as possible), together with the respective uncertainties for the chosen PDF set, the
        strong coupling $\alpha_s(M_Z)$  and the heavy quark masses $m_c$, $m_b$ and $m_t$. Once a PDF set is
        updated, the most recent version should be used.

    \item[2] {\textbf{Theory predictions for feasibility studies:}}
        Use any of the recent PDF sets. 

\end{enumerate} 

\section{Outlook}
Several aspects in PDF extraction have been briefly discussed here, which 
emphasize the importance of the selection of  consistent data sets and of the different theoretical 
assumptions in PDF fits. The main thrust
of the study has been the computation of benchmark cross sections for a variety of processes at
hadron colliders, including Higgs boson production in gluon-gluon fusion.
It has been illustrated
how different choices for the theoretical description of the hard scattering process and choices of
parameters have an impact on the predicted cross sections, and lead to systematic shifts that are
often significantly larger than the associated PDF and $\alpha_s(M_Z)$ uncertainties.
We aim at bringing the points considered and shortcomings exposed in the recent PDF4LHC recommendations~\cite{r8}
into discussions and to propose alternative recommendations for the PDF usage. 
Ideally, as a consequence of these discussions, the next recommendations would include these observations 
and would be published as a common work of the all PDF groups.
\\
\\
\textbf{Acknowledgments} I would like to thank authors of Ref.~\cite{r9}, A.~Accardi, S.~Alekhin, J.~Bl\" umlein, M.V.~Garzelli, K.~Lipka,
W.~Melnitchouk, S.~Moch, J.F.~Owens, E.~Reya, N.~Sato, A.~Vogt and O.~Zenaiev for their help in preparing the write-up.

\end{document}